\documentclass[11pt]{article}
\usepackage{times}
\usepackage{geometry}
\usepackage[utf8]{inputenc}
\usepackage{enumitem,amssymb}
\usepackage{graphicx}
\usepackage{fancyhdr}

\usepackage[authoryear]{natbib}
\setcitestyle{authoryear,open={(},close={)}}
\usepackage{mdframed} 

\mdfdefinestyle{theoremstyle}{
innertopmargin=\topskip,}
\mdtheorem[style=theoremstyle]{lrptextbox}{}

\begin{document}
\begin{center}
{\Large{\bf Canadian Astronomy on Maunakea: On Respecting Indigenous Rights}} \\
Hilding R.~Neilson \footnote{Department of Astronomy \& Astrophysics, University of Toronto, Toronto, Ontario, M5S~3H4, Canada. \\ e-mail:hilding.neilson@utoronto.ca}  \& Samantha Lawler \footnote{Campion College, University of Regina, Regina, SK S4S 0A2, Canada} 
\end{center} 
\noindent{\bf Executive Summary} \\
Canadian astronomy has, for decades, benefited from access to observatories and participating in international consortia on one of the best astronomical sites in the world: Maunakea. However, Maunakea is part of the unceded territory of the Native Hawaiian peoples and has always been of special significance to Hawaiian culture. The use of the summit and its science reserve has created tensions in the past decade, particularly with the development of the Thirty Meter Telescope. A meaningful and respectful response from the International astronomy community is still lacking. It is expected that the LRP 2020 will continue to support Canadian astronomy on Maunakea so a better official statement on the position and involvement of CASCA should be prepared. In this paper we present recommendations, based on the United Nation Declaration for the Rights of Indigenous Peoples, for the Canadian astronomical community to better support Indigenous rights on Maunakea and Hawaii  while providing clear guidelines for the astronomical community to participate in activities conducted on Indigenous land.  This framework is designed to motivate conversations with Indigenous communities regarding our place on Indigenous lands and our roles, and responsibilities toward the communities we are working with.  Furthermore, we propose this framework as a basis for engaging with communities around the world regarding consent for astronomical facilities. To this end, we propose six recommendations for CASCA:
\begin{enumerate}
\item Guaranteed funding from facilities for education and training for Native Hawaiians in relation to astronomy that includes significant outreach and community funding, telescope time for Native Hawaiian learners, scholarships for Native Hawaiians interested in astronomy in Canada;
\item Collaborate with institutions in Hawaii to develop training programs/courses for astronomers in Canada regarding  Hawaii, Native Hawaiian cultural traditions, history of colonization, and astronomy;
\item  Canada commits to maintaining a presence on Maunakea if and only if there is a clear and equitable agreement and consent for the Lease Renewal that benefits Native Hawaiians and clear, informed, and ongoing consent for the Maunakea Spectroscopic Explorer;
\item Canadian Support for Indigenous/Native Hawaiian usage and Indigenous/Native Hawaiian rights to the mountain;
\item That CASCA supports a process that requires clear Native Hawaiian consent for future projects, including the Maunakea Spectroscopic Explorer.  
\item Commitment that Canadian engagement on Maunakea must be consistent with the spirit of the Calls to Actions of the Truth and Reconciliation commission and the United Nations Declaration of the Rights of Indigenous Peoples.
\end{enumerate}



%


\section{Introduction}
Maunakea is one of the and arguably the best places in the world for optical and infrared astronomy.  Maunakea is also Native Hawaiian territory, Maunakea is Indigenous.  Astronomy has benefited from the colonization and annexation of Hawaii since the first telescopes were built on the Mountain.  Canadian astronomy has benefited through a number of direct collaborations such as the James Clerk Maxwell Telescope, the Canada-France-Hawaii Telescope, and the Gemini-North Telescope.  

However, the relationship Canadian astronomy has on Maunakea is also built on colonization. This result has been seen in the tension of the past decade with the Thirty-Metre Telescope and Native Hawaiians or Kanaka Maoli. At the time of submission of this paper, activists, the majority of which are Native Hawaiian, are camping at Maunakea in an act to elevate their voice and be heard by their government representatives and to protect the mountain from construction of TMT.  

We are presenting this paper with the goal of developing a set of guiding principles for continuing Canadian participation in facilities on Maunakea.  This includes the future of the Thirty-Metre Telescope, the expected negotiation of the lease renewal for facilities on the mountain and the  Maunakea Spectroscopic Explorer that is proposed to replace the Canada-France-Hawaii Telescope. There is the widespread perception that Native Hawaiians have received negligible benefit from the presence of telescopes on the mountain, while the astronomy community has benefited greatly. The Envision Maunakea report (http://envisionmaunakea.org) supports this feeling, especially in the context of TMT.

The Thirty-Metre Telescope is not the only controversy regarding astronomy on Maunakea. Almost twenty years ago, NASA and the Keck Telescopes proposed installing between four and six smaller outrigger telescopes to convert the Keck facility into an optical interferometer (https://www.skyandtelescope.com/astronomy-news/keck-outriggers-face-additional-roadblocks/).  The Office of Hawaiian Affairs opposed this in court based on a number of concerns of environmental and cultural impact on the mountain.  The project was canceled in 2006.  NASA had funded a preliminary study of the environmental and cultural impact of that project that included additional construction on a cinder cone considered sacred. That study was found to be insufficient because it only considered the impact on the specific area of construction.

The issues of TMT and the Keck Outriggers demonstrate complexities of consultation with Native Hawaiians, but also that we as a community are not adequately listening to the diversity of Native Hawaiian voices and respecting their wishes. In this work, we develop a series of recommendations for the Canadian astronomical community using the United Nations Declaration of the Rights of Indigenous Peoples as a framework.  The key element of this framework is that the requirement for consent from Indigenous Peoples as opposed to consultation.  Using these recommendations as a starting point for envisioning new facilities on Maunakea and in other territories of Indigenous peoples will help the Canadian Astronomical Community be more inclusion and move forward in a path of being less colonialist. This work is also designed to complement our other Community Paper E020: Indigenizing the next decade of astronomy in Canada.

\section{Astronomy and the UNDRIP}
The United Nations Declaration on the Rights of Indigenous Peoples was adopted on September 13, 2007\footnote{https://www.un.org/development/desa/indigenouspeoples/declaration-on-the-rights-of-indigenous-peoples.html} \citep{undrip}.  Only four nations voted against the resolution - Australia, Canada, New Zealand and the United States. The Canadian government issued a statement of support for the Declaration in 2010 and became a full supporter of it in 2016. The purpose of the UNDRIP is to resolve to affirm the rights of Indigenous people, to recognize the rights of Indigenous peoples on the lands where they are from, to recognize the need for reconciliation and to address historic injustices related to colonization and racism and to support Indigenous peoples working to organize and advocate for their rights.  The UNDRIP contains forty-six articles related to land use, language rights, anti-discrimination, etc.  The UNDRIP was adopted by Canada as part of the federal government's response to the Calls to Action of the Truth and Reconciliation Commission \citep{truth2015final}.

The Calls to Action of the Truth and Reconciliation Commission are ninety-six recommendations presented as a pathway for Canada to begin reconciliation with Indigenous peoples and to address its history of genocide against Indigenous people. In terms of astronomy, the Canadian astronomy community and Maunakea, there are four especially pertinent recommendations that our community can address:

\begin{itemize}
\item 13. We call upon the federal government to acknowledge that Aboriginal rights include Aboriginal language rights.
\item 47. We call upon federal, provincial, territorial, and municipal governments to repudiate concepts used to justify European sovereignty over Indigenous peoples and lands, such as the Doctrine of Discovery and terra nullius, and to reform those laws, government policies, and litigation strategies that continue to rely on such concepts.
\item 57. We call upon federal, provincial, territorial, and municipal governments to provide education to public servants on the history of Aboriginal peoples, including the history and legacy of residential schools, the United Nations Declaration on the Rights of Indigenous Peoples, Treaties and Aboriginal rights, Indigenous law, and Aboriginal--Crown relations. This will require skills-based training in intercultural competency, conflict resolution, human rights, and anti-racism.
\item 62. We call upon the federal, provincial, and territorial governments, in consultation and collaboration with Survivors, Aboriginal peoples, and educators, to: ii. Provide the necessary funding to post-secondary institutions to educate teachers on how to integrate Indigenous knowledge and teaching methods into classrooms. iii. Provide the necessary funding to Aboriginal schools to utilize Indigenous knowledge and teaching methods in classrooms.
\end{itemize}

The Calls to Action also include adopting the UNDRIP.  There are many recommendations that the Canadian community must consider including:
\begin{itemize}
\item Article 3 Indigenous peoples have the right to self-determination. By virtue of that right they freely determine their political status and freely pursue their economic, social and cultural development.
\item Article 8.2. States shall provide effective mechanisms for prevention of, and redress for: (a) Any action which has the aim or effect of depriving them of their integrity as distinct peoples, or of their cultural values or ethnic identities; (b) Any action which has the aim or effect of dispossessing them of their lands, territories or resources; (c) Any form of forced population transfer which has the aim or effect of violating or undermining any of their rights; (d) Any form of forced assimilation or integration; (e) Any form of propaganda designed to promote or incite racial or ethnic discrimination directed against them.
\item Article 11 1. Indigenous peoples have the right to practise and revitalize their cultural traditions and customs. This includes the right to maintain, protect and develop the past, present and future manifestations of their cultures, such as archaeological and historical sites, artefacts, designs, ceremonies, technologies and visual and performing arts and literature. 2. States shall provide redress through effective mechanisms, which may include restitution, developed in conjunction with indigenous peoples, with respect to their cultural, intellectual, religious and spiritual property taken without their free, prior and informed consent or in violation of their laws, traditions and customs.
\item Article 12 1. Indigenous peoples have the right to manifest, practise, develop and teach their spiritual and religious traditions, customs and ceremonies; the right to maintain, protect, and have access in privacy to their religious and cultural sites; the right to the use and control of their ceremonial objects; and the right to the repatriation of their human remains.
\item Article 18 Indigenous peoples have the right to participate in decision-making in matters which would affect their rights, through representatives chosen by themselves in accordance with their own procedures, as well as to maintain and develop their own indigenous decision-making institutions.
\item Article 19 States shall consult and cooperate in good faith with the indigenous peoples concerned through their own representative institutions in order to obtain their free, prior and informed consent before adopting and implementing legislative or administrative measures that may affect them.
\item Article 24 1. Indigenous peoples have the right to their traditional medicines and to maintain their health practices, including the conservation of their vital medicinal plants, animals and minerals. Indigenous individuals also have the right to access, without any discrimination, to all social and health services.
\item  Article 25 Indigenous peoples have the right to maintain and strengthen their distinctive spiritual relationship with their traditionally owned or otherwise occupied and used lands, territories, waters and coastal seas and other resources and to uphold their responsibilities to future generations in this regard.
\item Article 26 1. Indigenous peoples have the right to the lands, territories and resources which they have traditionally owned, occupied or otherwise used or acquired.
\item Article 28 1. Indigenous peoples have the right to redress, by means that can include restitution or, when this is not possible, just, fair and equitable compensation, for the lands, territories and resources which they have traditionally owned or otherwise occupied or used, and which have been confiscated, taken, occupied, used or damaged without their free, prior and informed consent. 2. Unless otherwise freely agreed upon by the peoples concerned, compensation shall take the form of lands, territories and resources equal in quality, size and legal status or of monetary compensation or other appropriate redress.
\item Article 29 1. Indigenous peoples have the right to the conservation and protection of the environment and the productive capacity of their lands or territories and resources. States shall establish and implement assistance programmes for indigenous peoples for such conservation and protection, without discrimination. 2. States shall take effective measures to ensure that no storage or disposal of hazardous materials shall take place in the lands or territories of indigenous peoples without their free, prior and informed consent.
\item Article 35 Indigenous peoples have the right to determine the responsibilities of individuals to their communities.
\end{itemize}

We interpret these articles as focusing on the rights of Indigenous peoples in terms of their territories; the rights of Indigenous peoples to self-govern and to practice their cultures and religions; and the right to restitution and reconciliation from colonizing parties. Canada and the Canadian astronomical community continue to benefit from colonization on Maunakea through the Maunakea lease agreement and through the TMT partnership, hence we have an ethical obligation to support any redress and to support the rights of Native Hawaiians.  Following these recommendations will impact our response to the expected lease renewal for Maunakea. 

\section{Connecting the recommendations of this work and the UNDRIP}
We propose six recommendations for CASCA to support which could form basis for our interactions with Maunakea and with Native Hawaiians to support their rights, culture, and beliefs as Indigenous people. 

{\bf Recommendation 1: Guaranteed funding from facilities for education and training for Native Hawaiians in relation to astronomy that includes significant outreach and community funding, telescope time for Native Hawaiian learners, scholarships for Native Hawaiians interested in astronomy in Canada.} Any project on Maunakea for which Canada is a partner needs to benefit and collaborate with Native Hawaiians both in terms of economic and educational support. The current Maunakea Lease agreement requires rent of one dollar per year and the University of Hawaii is given dedicated time on telescopes.  Given that most astronomers in the University of Hawaii system are not Native Hawaiian there is a clear lack of benefit to Native Hawaiians of having telescopes on the mountain.  

Any future discussions must include support for Native Hawaiians that offer economic and educational benefit in meaningful ways. In this way we propose that future projects commit a significant fraction of capital and operational costs supporting Native Hawaiians through occupational training and outreach to Hawaiian schools and communities.  This would include gifting some fraction of telescope time to Native Hawaiian schools and groups to practice and learn astronomy at the cutting edge of technology with support from observatory astronomers. Finally, this funding could also include scholarships and training for Native Hawaiian students to attend graduate schools in Hawaii or in Canada. These programs should operate independently of any one facility and they should not be solely dependent on obtaining consent for any one project on the mountain. Given our place on Maunakea currently, the Canadian astronomy community should work to implement and expand programs addressing this recommendation as soon as possible.

This recommendation addresses Articles 3, 28, and 35 of the UNDRIP as well as the 62nd Call to Action of the TRC.  The 62nd recommendation of the Calls to Action aims for educational funding at the post-secondary level.  By supporting Native Hawaiian education in astronomy (if students choose to do so) acts to help our community foster inclusion and respect for Native Hawaiian cultures. This addresses Article 3 by supporting Native Hawaiians rights to freely pursue economic development through these programs, and this program can act to redress the wrongs by the astronomy community on the mountain addressing Article 28.  Finally, these programs should be conducted in collaboration with Native Hawaiians, hence Native Hawaiians determine our responsibilities.

{\bf  Recommendation 2:  Collaborate with institutions in Hawaii to develop training programs/courses for astronomers in Canada regarding Hawaii, Native Hawaiian traditions, history of colonization, and astronomy.} One of the issues for astronomers working with observations from facilities on Maunakea is a lack of understanding of Hawaii, Native Hawaiians, and Native Hawaiian culture and history.  This issue is becoming more problematic in the era of queue observing in that the majority of people involved in using telescope data never see Hawaii or Maunakea. To be better partners and collaborators, Canadian astronomers should have access to training programs and courses in Native Hawaiian culture, history, and astronomy.  A collaborative development effort is essential.  Such an effort ensures that these courses would be supported by Native Hawaiian Kapuna and educators to help astronomers learn from the community while creating networks that support the educational initiatives outlived above. This also benefits the overall efforts in Canadian astronomy in that it offers more inclusive and diverse teachings than is typically seen here. 

This recommendation addresses the TRC Calls to Action recommendations 47, 57, and 62 and Article 35 of the UNDRIP.  By learning from Native Hawaiian Kapuna, astronomers will learn more about responsibilities in terms of Maunakea.  This recommendation addresses the Calls to Action by learning a non-western perspective of Hawaiian colonization and occupation, to develop new competencies in terms intercultural interactions and to integrate Indigenous knowledges into the astronomical field.

{\bf  Canada commits to maintaining a presence on Maunakea if and only if there is a clear and equitable agreement and consent for the Lease Renewal that benefits Native Hawaiians and clear, informed, and ongoing consent for the Maunakea Spectroscopic Explorer.} The recent experiences of the International astronomy community with respect to the proposed Thirty-Metre Telescope is that Indigenous consent matters and TMT currently lacks that support given the presence of activists on Maunakea against the TMT. As such the TMT project should work with the State of Hawaii and the Native Hawaiian community to determine if there is path forward or if TMT should withdraw.  This is necessary if the project is to continue in an ethical, non-violent manner.  Similarly, we need to consider our place on Maunakea in the next decade which includes the Master Lease renewal process.  The current lease does not offer any significant, explicit benefits for Native Hawaiians. If the Canadian community wishes to continue on Maunakea, in terms of Gemini and Canada-France-Hawaii Telescopes and in terms the proposed Maunakea Spectroscopic Explorer, then we should only proceed if there is an equitable agreement and clear consent from Native Hawaiians. 

The Canadian perspective regarding any equitable agreement should conform with the Articles of the UNDRIP.  As such this should support Hawaiian rights to access the mountain and supporting cultural and spiritual interactions with Maunakea as well as protection for the land and cultural icons.  This should support the rights of Hawaiians to be part of any decision making process related to astronomy on Maunakea.  In summary, any lease renewal should support and respect UNDRIP in its entirety and especially the Articles listed here if Canada is to be an equitable collaborator on the mountain.

{\bf  Canadian Support for Indigenous/Native Hawaiian usage and Indigenous/Native Hawaiian rights to the mountain.} The UNDRIP clearly notes the State should not impede the rights of Indigenous peoples to use their traditional territories for cultural, economic and, social benefit. In terms of Maunakea, this means that the presence of telescopes and facilities should not negatively impact access to the mountain, hence not contravene Hawaiian rights. This recommendation is supported by Articles 3, 11, 12, 19, 24, 25, 26, and 29.  Essentially, it is our duty as a community that benefits from access and use of land on Maunakea to remember that Maunakea is Native Hawaiian territory first and foremost.

{\bf  That CASCA supports a process that requires clear Native Hawaiian consent for future projects, including the Maunakea Spectroscopic Explorer.} In Canadian law, any project or construction on Indigenous territories requires a process of consultation.  The Thirty-Metre Telescope project also underwent a process of consultation with Native Hawaiian groups.  However, consultation is not consent. Any new and ongoing project on Maunakea, (or significant changes to facilities such as the example of the Keck Outriggers) requires Native Hawaiian consent.  Furthermore, the astronomy community does not have the privilege to define what consent is nor do we have the right to manufacture consent.  Consent can be given by Indigenous/Native Hawaiians and that consent can be rescinded and changed. Given the current climate in Hawaii and the presence of the Protectors camp on Maunakea, we can clearly see there are issues with consent in terms of the Thirty-Metre Telescope. 

{\bf  Commitment that Canadian engagement on Maunakea must be consistent with the Calls to Actions of the Truth and Reconciliation commission and the United Nations Declaration of the Rights of Indigenous Peoples;}
The Canadian government has endorsed the Calls to Action of the Truth and Reconciliation and has adopted the UNDRIP. These resolutions impact the relationships Canada has with Indigenous peoples everywhere, not just in Canada. As a community, CASCA has a duty to live up to the Calls to Action and the UNDRIP wherever possible.  This is especially important for facilities on Maunakea, in Chile, in Australia, and Indigenous lands everywhere.

\section{Summary and Conclusions}
These six recommendations are a first step for Canadian astronomy to be better and more inclusive partners on Indigenous lands.  These recommendations are meant to motivate a proper protocol for respecting Indigenous peoples and rights wherever astronomy is conducted by the Canadian community. We need to work to be better partners on Maunakea and we need to address our own historical injustices there, such as issues with the Maunakea Master Lease, the Keck Outriggers, and so on. 

While we wrote this document with respect to Maunakea and the rights of Native Hawaiians, but the UNDRIP and the Calls to Action are for relationships with Indigenous peoples everywhere.  The Canadian community is already participating in facilities in Chile such as ALMA and Gemini, CHIME/DRAO and DAO in British Columbia, and the Mount Megantic Telescope in Quebec, and this involvement is expected to continue or even expand over the next decade in these locations and also potentially the Square Kilometre Array in Australia and South Africa.  The success of Canadian astronomy has been built on Indigenous lands and will continue to depend on facilities built on Indigenous lands for decades to come.  We, as a community, have an ethical duty to engage and listen to Indigenous peoples. 

This community paper has been written by astronomers, none of whom are Native Hawaiian.  One authors is Indigenous, but not Native Hawaiian.  As such this makes this work incomplete as it lacks Native Hawaiian voices. This is another reason why we should consider these recommendations as first steps for the Canadian community to move forward with.




\begin{lrptextbox}[How does the proposed initiative result in fundamental or transformational advances in our understanding of the Universe?]
The initiatives are important to improve inclusivity and equity in astronomy and to learn from Hawaiian and Indigenous peoples.

\end{lrptextbox}

\begin{lrptextbox}[What are the main scientific risks and how will they be mitigated?]
N/A

\end{lrptextbox}

\begin{lrptextbox}[Is there the expectation of and capacity for Canadian scientific, technical or strategic leadership?] 
The recommendations require and depend on leadership from the Canadian community to develop the proposed initiatives.
\end{lrptextbox}

\begin{lrptextbox}[Is there support from, involvement from, and coordination within the relevant Canadian community and more broadly?] 
There is engagement in this from the Canadian community on Maunakea and across Canada.
\end{lrptextbox}

\begin{lrptextbox}[Will this program position Canadian astronomy for future opportunities and returns in 2020-2030 or beyond 2030?] 
N/A
\end{lrptextbox}

\begin{lrptextbox}[In what ways is the cost-benefit ratio, including existing investments and future operating costs, favourable?] 
N/A
\end{lrptextbox}

\begin{lrptextbox}[What are the main programmatic risks
and how will they be mitigated?] 
N/A
\end{lrptextbox}

\begin{lrptextbox}[Does the proposed initiative offer specific tangible benefits to Canadians, including but not limited to interdisciplinary research, industry opportunities, HQP training,
EDI,
outreach or education?] 
The initiatives in this work focus on being inclusive of Indigenous knowledge and respecting Indigenous rights. These initiatives support HQP training in inclusive ways and will contribute to diversity in the Canadian Community.
\end{lrptextbox}

\bibliography{WPMK-refs} 

\end{document}